# First-principles Study of Structural and Electronic Properties of Mn-doped $Cu_2NiXY_4$ (X=Sn, Ge, Si; Y=S, Se) Chalcogenide Semiconductors

I.M. Raufzoda*, D.D. Nematov and A.S. Burhonzoda

*S.U. Umarov Physical-Technical Institute of the National Academy of Sciences of Tajikistan*

**Abstract**: In this work, the effect of partial substitution of Mn by Ni on the structural and electronic properties of kesterite systems $Cu_2NiXY_4$ (X = Sn, Ge, Si; Y = S, Se) was studied using density functional theory (DFT). The mBJ+U method was used to characterize the bandgap more accurately. To the best of our knowledge, a systematic comparative study of Mn substitution in the entire $Cu_2NiXY_4$ (X = Sn, Ge, Si; Y = S, Se) family has not been previously performed. Due to crystallographic constraints of the kesterite unit cell, the substitution of a single Ni atom with Mn corresponds to 50%. This configuration was chosen to study the effect of Mn substitution on the structural and electronic properties of $Cu_2NiXY_4$ (X = Sn, Ge, Si; Y = S, Se) compounds, whereas the study of lower concentrations requires the use of a supercell and is the subject of further research. The calculation results show that the partial substitution of Ni with Mn preserves the tetragonal structure of kesterite and significantly alters the electronic structure. In all the compounds studied, the bandgap decreases from 1.028–3.397 to 1.007–3.333 eV. For example, in the $Cu_2NiSnS_4$ system, the bandgap width decreases from 1.59 eV to 1.49 eV. The narrowing of the bandgap results from hybridization between the Mn-3d, Cu-3d, and S/Se-p orbitals near the band edges, leading to a redistribution of electronic states around the Fermi level. The results demonstrate that Mn substitution is an effective strategy for controlling the electronic properties of $Cu_2NiXY_4$ kesterites, offering great promise for use in optoelectronic devices where adjustable bandgaps are required.

## 1. INTRODUCTION

Solar energy is one of the most important renewable energy sources, and its conversion into electrical energy using photovoltaic devices has attracted the attention of scientists due to its efficiency and low manufacturing cost [1]. In this process, the light-absorbing layer plays a key role, and nanoconductive kesterites have been proposed as promising materials for use in thin-film photovoltaic cells [2, 3]. Kesterite $Cu_2ZnSnS_4$ (CZTS) is considered one of the most promising materials due to its abundance of constituent elements, non-toxicity, high light absorption coefficient, and relatively inexpensive synthesis technology [4]. Theoretical calculations predict an energy conversion efficiency of up to 32.2% [5], while the experimentally achieved efficiency is 12.6% [6], indicating the need to improve its properties.

Previous theoretical and experimental studies have shown that replacing Zn with Ni significantly changes the electronic structure, improves optical absorption, and increases the potential of kesterite materials for photovoltaic applications [7]. One such compound is $Cu_2NiGeS_4$ (CNGS), which shows good light absorption and potential for use in solar cells. It is also used to improve the efficiency of the $MoS_2$ interlayer [8].

*Address correspondence to these authors at the S.U. Umarov Physical-Technical Institute of the National Academy of Sciences of Tajikistan;
E-mail: E-mail: raufov.i@mail.ru

Despite significant progress in studying the structural, electronic, and optical properties of $Cu_2NiXY_4$ (X = Sn, Ge, Si; Y = S, Se), information on the effects of transition metal substitution on these materials is still limited [9-14]. In particular, no study has been reported on Mn substitution in all $Cu_2NiXY_4$ (X = Sn, Ge, Si; Y = S, Se) compounds. Transition metal substitution significantly changes the electronic structure through d-orbital hybridization, which provides an effective way to design the band gap and improve the functional properties for use in photovoltaic and optoelectronic devices. Therefore, information on the electronic structure changes induced by Mn is of great scientific and technical importance.

Among the transition metals, Mn is attractive due to its half-filled 3d electronic configuration, which facilitates strong d-orbital interactions while maintaining the semiconducting nature of the material's lattice. Unlike Fe, Co, and Cr, Mn often induces moderate lattice distortions along with effective modulation of the electronic structure, making it a promising candidate for band gap engineering in photovoltaic and optoelectronic semiconductors.

In this study, we systematically investigated the partial substitution of Ni by Mn in $Cu_2NiXY_4$ (X = Sn, Ge, Si; Y = S, Se) systems. The electronic structure optimization was performed using the SCAN functional, while the electronic properties were calculated using the mBJ+U method to obtain reliable band gap predictions. This study provides a unique comparison of the structural deformation, band gap changes, and density of states distributions induced by Mn substitution.

## 2. METHOD AND METHODOLOGY

Ab initio calculations were performed using density functional theory (DFT) using the VASP program [15]. To characterize the electronic interactions, the meta-GGA SCAN functional [16] was used for geometric optimization. Integration tests were performed to determine the optimal cutoff energy of the plane wave (600 eV) and the Monhost-Pak k-point grid (6x6x3). The convergence of the total energy to the level of $10^{-5}$ eV and the residual forces on each atom were less than 0.01 eV/Å was ensured. Electronic structure calculations were performed using the mBJ+U functional. The Hubbard U correction has been used to improve the characterization of three-dimensional localized electrons in Cu, Ni, and Mn, for which semilocal exchange correlation functions often underestimate electron localization due to self-interference errors. In this study, the value of U=5 for Cu, Ni, and Mn was adopted based on published theoretical studies, where similar values provide reliable electronic structure and bandgap predictions [17-21]. The Hubbard correction mainly affects the localization of three-dimensional transition metal states and their hybridization with S/Se-p orbitals. As a result, the calculated valence and conduction band positions become more physically realistic, leading to better agreement with experimental results. The reliability of the chosen U parameters is confirmed by the good agreement between the calculated bandgaps of the pure compounds and the available experimental values.

All calculations were performed using PAW pseudopotentials, whose valence configuration is as follows: Cu: $3d^{10}$ $4s^1$, Ni: $3d^8$ $4s^2$, Mn: $3p^6$ $3d^6$ $4s^1$, Sn: $5s^2$ $5p^2$, Ge: $4s^2$ $4p^2$, Si: $3s^2$ $3p^2$, S: $3s^2$ $3p^4$, Se: $4s^2$ $4p^4$.

There are two Ni atoms in the primary structure of $Cu_2NiXY_4$ (X = Sn, Ge, Si; Y = S, Se) kesterite. Replacing one of these Ni atoms with Mn corresponds to a 50% substitution level. This concentration represents the minimum substitution that can be investigated without constructing a supercell and therefore provides an efficient computational model for evaluating the intrinsic effects of Mn on the crystal and electronic structure. The aim of this study is to identify the main electronic and structural trends induced by Mn substitution, rather than a specific experimental composition. Using the same substitution level for all investigated compounds also allowed for a reasonable comparison between all $Cu_2NiXY_4$ (X = Sn, Ge, Si; Y = S, Se) compounds. It should be noted that experimentally obtained Mn concentrations are often less than 50%. Investigating Mn atom substitution requires a larger supercell along with sampling multiple atomic configurations, which significantly increases the computational cost. Such concentration-dependent studies, including magnetic ordering and defect formation, will be the subject of future research.

## 3. RESULTS AND DISCUSSION

To analyze the effect of partial substitution of Ni with $Mn^{2+}$ atoms on the kesterite structure of $Cu_2NiXY_4$ (X = Sn, Ge, Si; Y = S, Se), a full geometry optimization was performed using the meta-GGA SCAN functional. Figure (**1**) shows the optimized structure, in which the Mn substitution preserves the basic tetragonal kesterite structure and does not cause significant changes in symmetry.

Analysis of the local structure of the Mn atom shows that the substitution of Ni with Mn leads to the formation of a deformed tetrahedral coordination environment with S and Se anions (Fig. **1**). The Mn-Y bond distances vary depending on the cationic and anionic composition.

The crystal structures were modeled using the VESTA software package [22]. When changing from Sn to Ge and Si, a systematic decrease in the Mn-Y distances is observed. For example, in the $Cu_2NiSnS_4$ compound, the Mn-S distance is 2.39 Å, and when changing to the $Cu_2NiGeS_4$ and $Cu_2NiSiS_4$ systems, the Mn-S distance is 2.37 and 2.38 Å, respectively. In the $Cu_2NiSnSe_4$ system, the Mn-Se distance is 2.51 Å, and in the $Cu_2NiGeSe_4$ and $Cu_2NiSiSe_4$ systems, it is 2.50 and 2.49 Å, respectively. The results showed that when replacing the anions from S to Se, the Mn-Se bond distances increase relative to Mn-S, which is mainly explained by the larger atomic radius of Se.

This geometric change indicates that the local environment of Mn is significantly sensitive to the chemical composition of the system, leading to further changes in the electronic structure.

A systematic comparison of all the $Cu_2NiXY_4$ (X = Sn, Ge, Si; Y = S, Se) compounds substituted with Mn reveals several clear composition-dependent trends. When the group IV cation Sn is replaced by Ge and Si, the lattice parameter and its volume gradually decrease due to the smaller ionic radii of Ge and Si. Conversely, when sulfur is replaced by selenium, the lattice becomes larger due to the larger ionic radii of selenium. For example, the $Cu_2Ni_{0.5}Mn_{0.5}SnS_4$ system has lattice parameters a=b (5.407 Å) and c (10.692 Å), while when going to the $Cu_2Ni_{0.5}Mn_{0.5}SnSe_4$ system they have values of a=b (5.700 Å) and c (11.200 Å). This structural change directly affects the local environment around Mn and consequently changes the electronic structure. A similar systematic behavior is observed for the electronic properties. Regardless of the cation or anion composition, Mn substitution consistently leads to a narrowing of the band gap. Although the band gap narrowing varies slightly among the studied compounds, the general trend remains the same for all Å compounds. This behavior indicates that the electronic effect of Mn substitution is strong and only slightly depends on the chemical composition of the compound. DOS calculations further confirm this systematic trend. In all compounds, the 3d states derived from Mn appear near the band edges and hybridize with the Cu-3d and S/Se-p orbitals, resulting in a similar distribution of electronic states, despite the differences in chemical composition. Therefore, Mn substitution acts as an effective way to tune the electronic structure of $Cu_2NiXY_4$ (X = Sn, Ge, Si; Y = S, Se) kesterites.

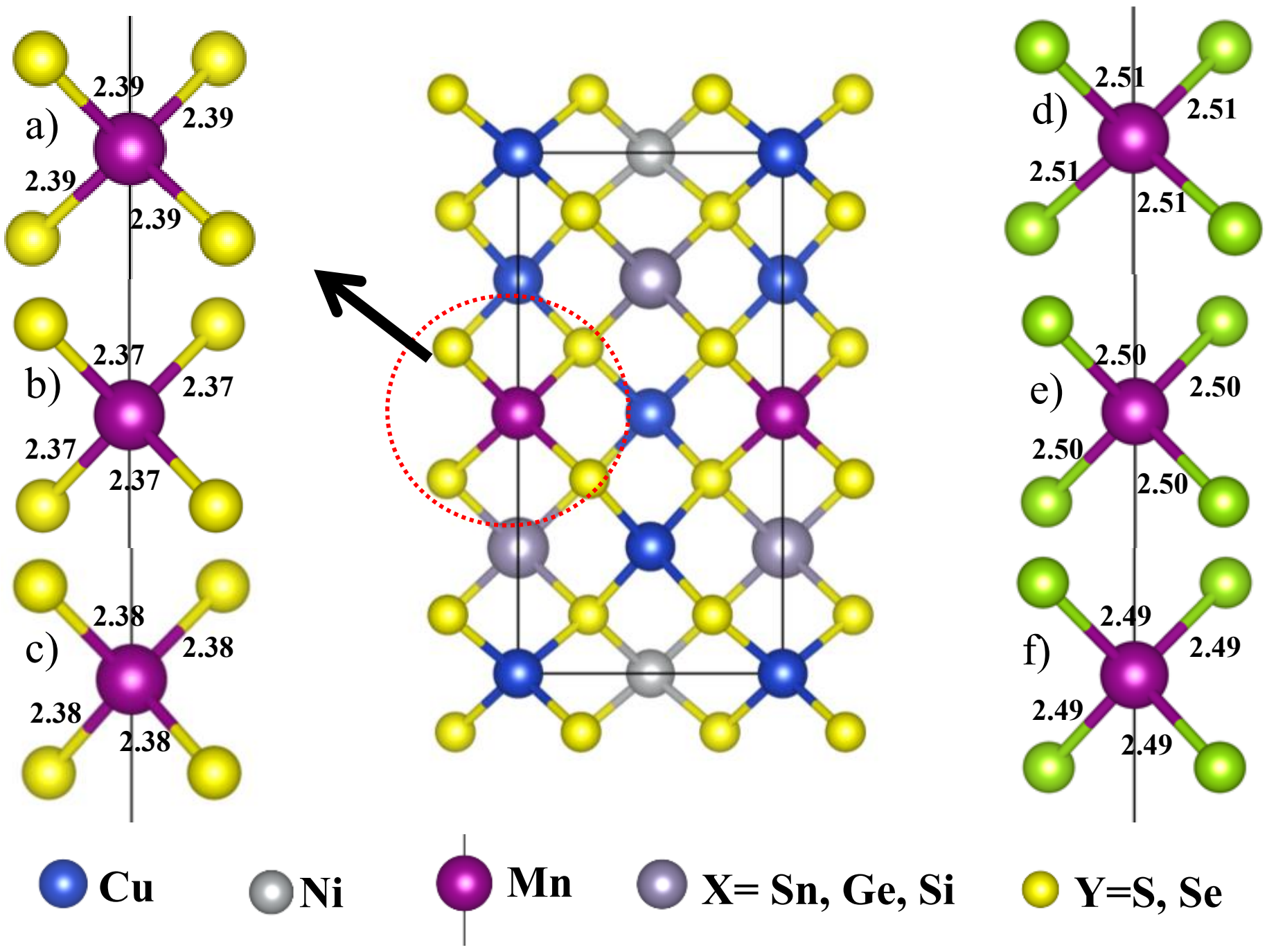

**Figure 1:** Crystal structure of $Cu_2NiXY_4$ (X = Sn, Ge, Si; Y = S, Se) with partial substitution of Ni by $Mn^{2+}$ and local coordination environment of Mn in the systems (a) - $Cu_2NiSnS_4$, (b) - $Cu_2NiGeS_4$, (c) - $Cu_2NiSiS_4$, (d) - $Cu_2NiSnSe_4$, (e) - $Cu_2NiGeSe_4$, (f) - $Cu_2NiSiSe_4$.

**Table 1: Structural Parameters of $Cu_2NiXY_4$ (X = Sn, Ge, Si; Y = S, Se) with Partial Substitution of Ni by $Mn^{2+}$, Calculated Using the SCAN Functional**

| System | Lattice Constant | | | |
|---|---|---|---|---|
| | a=b Å | c Å | α = β = γ | V Å³ |
| $Cu_2Ni_{0.5}Mn_{0.5}SnS_4$ | 5.407 | 10.692 | 90 | 312.588 |
| $Cu_2Ni_{0.5}Mn_{0.5}GeS_4$ | 5.233 | 10.676 | 90 | 292.355 |
| $Cu_2Ni_{0.5}Mn_{0.5}SiS_4$ | 5.181 | 10.557 | 90 | 283.379 |
| $Cu_2Ni_{0.5}Mn_{0.5}SnSe_4$ | 5.700 | 11.200 | 90 | 363.888 |
| $Cu_2Ni_{0.5}Mn_{0.5}GeSe_4$ | 5.521 | 11.280 | 90 | 343.831 |
| $Cu_2Ni_{0.5}Mn_{0.5}SiSe_4$ | 5.481 | 11.115 | 90 | 333.910 |

Table **2** shows a comparison of the bandgap of the systems $Cu_2NiXY_4$ (X = Sn, Ge, Si; Y = S, Se) before and after the substitution of Ni by Mn. The calculated band gap values of the $Cu_2NiXY_4$ (X = Sn, Ge, Si; Y = S, Se) compounds show good agreement with experimental values and theoretical calculations, demonstrating the reliability of this calculation approach. For example, the calculated band gap of $Cu_2NiSnS_4$ of 1.592 eV is in good agreement with the experimental value of 1.6 eV and the previously reported theoretical value of 1.67 eV. Similarly, the calculated band gap of $Cu_2NiGeS_4$ of 2.039 eV is in good agreement with the experimental value of 1.8 eV, and the calculated band gap of $Cu_2NiSnSe_4$ of 1.028 eV is in good agreement with the reported experimental value of 1.1 eV. It is expected that the small difference between the calculated and experimental values depends on the synthesis conditions, crystal quality, secondary phase defects, temperature, and

**Table 2: Band Gap (Eg) of $Cu_2NiXY_4$ (X = Sn, Ge, Si; Y = S, Se) with Partial Substitution of Ni by $Mn^{2+}$ Calculated Using the mBJ+U Approach**

| System | This Work | Literature | |
|---|---|---|---|
| | mBJ+U | Exp | Calc |
| $Cu_2NiSnS_4$ | 1.592 | 1.6 [23] | 1.67 [24] |
| $Cu_2Ni0_{.5}Mn_{0.5}SnS_4$ | 1.487 | | |
| $Cu_2NiGeS_4$ | 2.039 | 1.8 [25] | 1.6-1.75 [25] |
| $Cu_2Ni_{0.5}Mn_{0.5}GeS_4$ | 1.981 | | |
| $Cu_2NiSiS_4$ | 3.397 | - | - |
| $Cu_2Ni_{0.5}Mn_{0.5}SiS_4$ | 3.333 | | |
| $Cu_2NiSnSe_4$ | 1.028 | 1.1 [26] | - |
| $Cu_2Ni_{0.5}Mn_{0.5}SnSe_4$ | 1.007 | | |
| $Cu_2NiGeSe_4$ | 1.266 | | |
| $Cu_2Ni_{0.5}Mn_{0.5}GeSe_4$ | 1.188 | | |
| $Cu_2NiSiSe_4$ | 2.402 | | |
| $Cu_2Ni_{0.5}Mn_{0.5}SiSe_4$ | 2.328 | | |

measurement method, while the present calculation describes the crystal structure without defects. In addition, This means that these differences between theoretical studies mainly arise from the choice of exchange-correlation functional parameters and Hubbard U. To our knowledge, no experimental data or theoretical calculations have been reported for $Cu_2NiXY_4$ (X = Sn, Ge, Si; Y = S, Se) compounds in which the Ni atom is partially replaced by Mn. Therefore, the present results constitute the first systematic set of theoretical data for these materials and provide useful guidance for future synthesis and experimental characterization.

Comparative analysis of the density of electronic states allows us to evaluate the potential of these compounds as active photovoltaic layers and materials for heterostructures. To determine the contribution of elements to the formation of the valence and conduction bands, the density of electronic states (DOS) for the systems $Cu_2NiXY_4$ (X = Sn, Ge, Si; Y = S, Se) with partial substitution of Ni by $Mn^{2+}$ atoms is shown in Fig. (**2**). DOS analysis provides important insights into the microscopic origin of the electronic structure changes induced by Mn substitution. In the initial compounds, the valence maxima consist mainly of Cu-3d and S/Se-p states, while the conduction minima consist of antibonding states involving group IV cations and S/Se atoms. After Mn substitution, the local Mn-3d states interact strongly with these band edge states, leading to increased d-p hybridization and a redistribution of the electron state density near the Fermi level. This hybridization reduces the energy between the valence maxima and the conduction band minimum and leads to a narrowing of the band gap in all the compounds studied. Since the additional Mn-derived states remain near the band edges rather than occupying discrete deep levels within the band gap, the semiconducting nature of the initial structure is preserved. From a device perspective, narrowing the bandgap can improve visible light absorption and facilitate the generation of photoexcited carriers, both of which are useful for photovoltaic applications. In addition, the increased density of electronic states near the band edges can affect the excitation and transport of carriers through altered orbital interactions. However, estimating the amount of carrier mobility, effective masses, recombination mechanisms, and carrier lifetimes require specialized transport calculations that are beyond the scope of this study.

From the perspective of photovoltaic applications, reducing the band gap leads to improved visible light absorption. The calculated band gap values of $Cu_2NiSnS_4$, $Cu_2NiGeS_4$, $Cu_2NiSnSe_4$ and $Cu_2NiGeSe_4$ compounds, in which Mn is partially replaced by Ni, remain within the optimal range (1.0-2.0), which is generally considered suitable for thin-film solar absorbers. As a result, Mn substitution provides a practical way to tune the optical absorption without structural instability.

In addition to its application in photovoltaic devices, the modification of the electronic structure by Mn substitution can also be useful for optoelectronic devices, photodetectors, and semiconductor heterostructures in which the band gap is tunable. The studied compounds retain the tetragonal kesterite structure after Mn substitution, providing the structural stability required for practical device fabrication. Furthermore, $Cu_2NiXY_4$ (X = Sn, Ge, Si; Y = S, Se) compounds are mainly composed of abundant earth elements. They are an attractive alternative to traditional semiconductor doping materials that rely on rare or toxic elements. Therefore, Mn substitution is an effective way to tailor the functional properties of environmentally friendly energy materials.

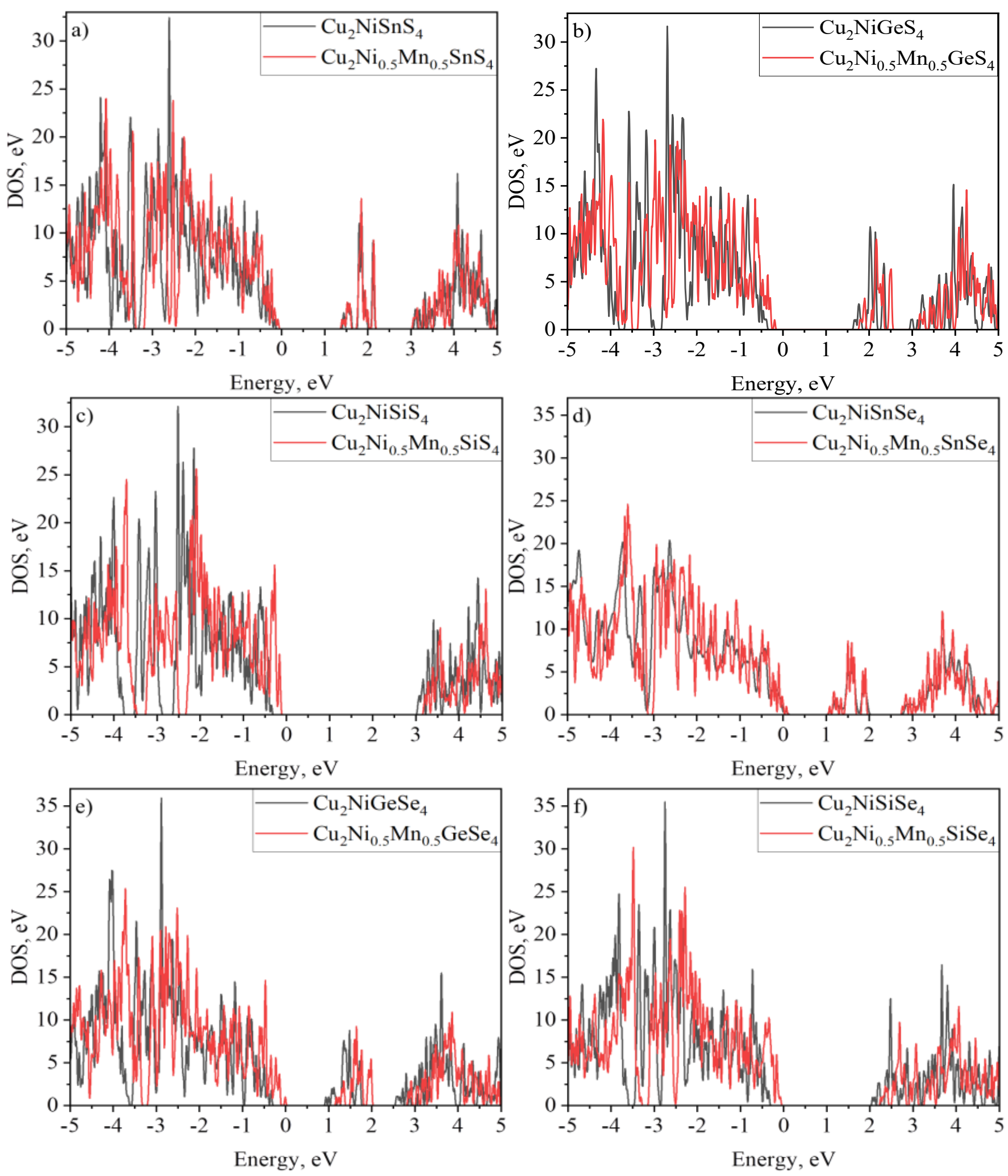

**Figure 2:** Total density of states (DOS) of $Cu_2NiXY_4$ (X = Sn, Ge, Si; Y = S, Se) with partial substitution of Ni by $Mn^{2+}$: a) $Cu_2NiSnS_4$ b) $Cu_2NiGeS_4$ c) $Cu_2NiSiS_4$ d) $Cu_2NiSnSe_4$ e) $Cu_2NiGeSe_4$ f) $Cu_2NiSiSe_4$.

## 4. CONCLUSION

In this study, the effect of partial substitution of Ni atoms by Mn on the structural and electronic properties of $Cu_2NiXY_4$ kesterites (X = Sn, Ge, Si; Y = S, Se) was investigated using the density functional theory (DFT) method. Geometric optimization calculations showed that after Mn substitution, the tetragonal structure of kesterite is preserved and deformation occurs throughout the structure. The change in lattice parameters is associated with the difference in ionic radii and chemical bonding properties of the constituent elements.

Electronic structure calculations show that Mn substitution leads to a narrowing of the band gap width in all studied compounds. This behavior is due to the formation of Mn-3d electronic states and their hybridization with Cu-3d orbitals and S/Se-p anions near the band edges, which leads to a narrowing of the band gap width and a redistribution of the density of electronic states. These changes demonstrate that Mn substitution is an effective way to engineer the electronic structure and tune the bandgap width in $Cu_2NiXY_4$ (X = Sn, Ge, Si; Y = S, Se) compounds. The calculated bandgap values remain within the range of interest for photovoltaic and optoelectronic applications, indicating that substituted $Cu_2NiXY_4$ (X = Sn, Ge, Si; Y = S, Se) compounds represent promising earth-abundant semiconductor materials for next-generation thin-film panels. However, the present study is limited to a 50% Mn substitution in the initial cell. The low Mn concentration, magnetic properties, intrinsic defects, carrier transport properties, and the effects of the final temperature were not considered and need to be investigated using a larger supercell. Future studies will focus on the concentration-dependent Mn substitution, magnetic and optical properties, and defect engineering to confirm the theoretical predictions reported in this study.

## CONFLICT OF INTEREST

The authors declare that there are no known competing financial interests or personal relationships that could have appeared to influence the work reported in this paper.

## ACKNOWLEDGEMENTS

The authors acknowledge the support of the International Science and Technology Center (ISTC) under Project TJ-0040. Special thanks are extended to the Government of Japan for its financial contribution and for supporting the establishment of a modern research laboratory at the S.U. Umarov Physical–Technical Institute of the National Academy of Sciences of Tajikistan.

This research was also supported by the International Innovation Center for Nanotechnologies of the CIS and the Interstate Fund for Humanitarian Cooperation of the CIS Member States, and it aligns with their strategic interests (grant No. 22-112).

## USE OF AI TOOLS

Artificial intelligence tools were used solely for language polishing, grammatical correction, and minor editorial assistance. All scientific content, data analysis, and interpretation of the results were performed exclusively by the authors. The authors carefully reviewed and verified all AI-assisted text to ensure accuracy, originality, and compliance with ethical publishing standards, including the COPE guidelines.’

## AUTHORSHIP CONTRIBUTION STATEMENT

I. Raufzoda: Resources, Investigation, Formal analysis;

D. Nematov: Writing – Original draft & review, Supervision, Project administration;

A. Burhonzoda: Methodology, Data curation, review & editing.